\documentclass[12pt,preprint]{aastex}
\usepackage{color}

\slugcomment{Accepted on 2014 July 21 for publication in ApJ}

\shorttitle{Optical Properties of (162173) 1999 JU3}
\shortauthors{Ishiguro et al.}

\begin{document}

\title{Optical Properties of (162173) 1999 JU3:\\
In Preparation for the JAXA Hayabusa 2 Sample Return Mission\footnote{This
work was conducted as part of the activities of the JAXA Hayabusa 2 Ground-Based
Observation Sub-Group.}}

\author{Masateru \textsc{Ishiguro}\altaffilmark{\dag}}
\affil{Department of Physics and Astronomy, Seoul National University,\\
Gwanak, Seoul 151-742, South Korea}

\author{Daisuke \textsc{Kuroda}}
\affil{Okayama Astrophysical Observatory, National Astronomical Observatory of Japan, \\
Asaguchi, Okayama 719-0232, Japan}

\author{Sunao \textsc{Hasegawa}}
\affil{Institute of Space and Astronautical Science, Japan Aerospace Exploration Agency,\\
Sagamihara, Kanagawa 252-5210, Japan}

\author{Myung-Jin \textsc{Kim}\altaffilmark{\ddag}}
\affil{Department of Astronomy, Yonsei University, \\
50 Yonsei-ro, Seodaemun-gu, Seoul 120-749, South Korea}

\author{Young-Jun \textsc{Choi}}
\affil{Korea Astronomy and Space Science Institute,\\
776 Daedeokdae-ro, Yuseong-gu, Daejeon 305-348, South Korea}

\author{Nicholas \textsc{Moskovitz}}
\affil{Lowell Observatory, 1400 W. Mars Hill Rd., Flagstaff, AZ 86001, USA}

\author{Shinsuke \textsc{Abe}}
\affil{Department of Aerospace Engineering, Nihon University,\\
7-24-1 Narashinodai Funabashi, Chiba 274-8501, Japan}

\author{Kang-Sian \textsc{Pan}}
\affil{Institute of Astronomy, National Central University, \\
300 Jhongda Road, Jhongli, Taoyuan 32001, Taiwan}

\author{Jun \textsc{Takahashi}, Yuhei \textsc{Takagi}, Akira \textsc{Arai}}
\affil{Nishi-Harima Astronomical Observatory, Center for Astronomy,
University of Hyogo, \\Sayo, Hyogo 679-5313, Japan}

\author{Noritaka \textsc{Tokimasa}}
\affil{Sayo Town Office, 2611-1 Sayo, Sayo-cho, Sayo, Hyogo 679-5380, Japan}

\author{Henry H. \textsc{Hsieh}}
\affil{Academia Sinica Institute of Astronomy and Astrophysics, \\
Roosevelt Rd., Taipei 10617, Taiwan}

\author{Joanna E. \textsc{Thomas-Osip}, David J. \textsc{Osip}}
\affil{The Observatories of the Carnegie Institute of Washington, Las
Campanas Observatory, \\Colina El Pino, Casilla 601, La Serena, Chile}

\author{Masanao \textsc{Abe}, Makoto \textsc{Yoshikawa}
\affil{Institute of Space and Astronautical Science, Japan Aerospace Exploration Agency,\\
Sagamihara, Kanagawa 252-5210, Japan}
}

\author{Seitaro \textsc{Urakawa}}
\affil{Bisei Spaceguard Center, Japan Spaceguard Association, \\
1716-3 Okura, Bisei-cho, Ibara, Okayama 714-1411, Japan}

\author{Hidekazu \textsc{Hanayama}}
\affil{Ishigakijima Astronomical Observatory, National Astronomical
Observatory of Japan, \\
1024-1 Arakawa, Ishigaki, Okinawa 907-0024, Japan}

\author{Tomohiko \textsc{Sekiguchi}}
\affil{Department of Teacher Training, Hokkaido University of Education, \\
9 Hokumon, Asahikawa 070-8621, Japan}

\author{Kohei \textsc{Wada}, Takahiro \textsc{Sumi}}
\affil{Department of Earth and Space Science, Graduate School of Science, Osaka University,
1-1 Machikaneyama, Toyonaka, Osaka 560-0043, Japan}

\author{Paul J. \textsc{Tristram}}
\affil{Mount John Observatory, P.O. Box 56, Lake Tekapo 8770, New Zealand}

\author{Kei \textsc{Furusawa}, Fumio \textsc{Abe}}
\affil{Solar-Terrestrial Environment Laboratory, Nagoya University, Nagoya, 464-8601, Japan}

\author{Akihiko \textsc{Fukui}}
\affil{Okayama Astrophysical Observatory, National Astronomical
Observatory of Japan, Asaguchi, Okayama 719-0232, Japan}

\author{Takahiro \textsc{Nagayama}}
\affil{Nagoya University, Furo-cho, Chikusa-ku, Nagoya, Aichi 464-8602,
Japan}

\author{Dhanraj S. \textsc{Warjurkar}}
\affil{Department of Physics and Astronomy, Seoul National University,\\
Gwanak, Seoul 151-742, South Korea}

\author{Arne \textsc{Rau}, Jochen \textsc{Greiner}, Patricia \textsc{Schady}, Fabian \textsc{Knust}}
\affil{Max-Planck-Institut f{\"u}r extraterrestrische Physik,\\
Giessenbachstra{\ss}e, Postfach 1312, 85741, Garching, Germany}

\author{Fumihiko \textsc{Usui}}
\affil{Department of Astronomy, Graduate School of Science,
The University of Tokyo,\\ 7-3-1 Hongo, Bunkyo-ku, Tokyo 113-0033, Japan}

\author{Thomas G. \textsc{M{\"u}ller}}
\affil{Max-Planck-Institut f{\"u}r extraterrestrische Physik,\\
Giessenbachstra{\ss}e, Postfach 1312, 85741, Garching, Germany}

\altaffiltext{\dag}{Visiting Scientist, Institut de mecanique celeste et de calcul des ephemerides,
Observatoire de Paris, 77 Avenue Denfert Rochereau, F-75014 Paris, France}

\altaffiltext{\ddag}{Korea Astronomy and Space Science Institute,
776 Daedeokdae-ro, Yuseong-gu, Daejeon 305-348, South Korea}

\begin{abstract}
We investigated the magnitude--phase relation of (162173) 1999 JU3, a target
asteroid for the JAXA Hayabusa 2 sample return mission.
We initially employed the international Astronomical Union's $H$--$G$ formalism
but found that it fits less well using a single set of parameters. To improve the inadequate fit,
we employed two photometric functions, the Shevchenko and Hapke functions.
With the Shevchenko function, we found that the magnitude--phase
relation exhibits linear behavior in a wide phase angle range ($\alpha$ = 5--75\arcdeg) and shows weak
nonlinear opposition brightening at $\alpha<$ 5\arcdeg, providing a more reliable absolute magnitude of
$H_\mathrm{V}$ = 19.25 $\pm$ 0.03. The phase slope (0.039 $\pm$ 0.001 mag deg$^{-1}$) and 
opposition effect amplitude (parameterized by the ratio of intensity at $\alpha$=0.3\arcdeg\
to that at $\alpha$=5\arcdeg, $I(0.3\arcdeg)$/$I(5\arcdeg)$=1.31$\pm$0.05)
are consistent with those
of typical C-type asteroids. We also attempted to determine the parameters for the Hapke model,
which are applicable for constructing the surface
reflectance map with the Hayabusa 2 onboard cameras. Although we could not constrain the full set of
Hapke parameters, we obtained possible values, $w$=0.041, $g$=-0.38, $B_0$=1.43, and $h$=0.050,
assuming a surface roughness parameter $\bar{\theta}$=20\arcdeg.
By combining our photometric study with a thermal model of the asteroid (M{\"u}ller et al.
in preparation), we obtained a geometric albedo of $p_\mathrm{v}$ = 0.047 $\pm$ 0.003, phase
integral $q$ = 0.32 $\pm$ 0.03, and Bond albedo $A_\mathrm{B}$ = 0.014 $\pm$ 0.002,
which are commensurate with the values for common C-type asteroids.
\end{abstract}


\section{Introduction}

A preliminary survey of mission target body is important to implement the space project
smoothly and safely. Hayabusa 2 is the successor to the original Hayabusa mission, which was
concluded with great success in obtaining samples from sub-km S-type asteroid (25143) Itokawa (1998 SF36).
A near-Earth asteroid, (162173) 1999 JU3 (hereafter JU3), is a primary target asteroid for
the Hayabusa 2. The spacecraft is scheduled to launch in late 2014, arrive at JU3 in 2018,
and return to Earth with the asteroidal sample around late 2020. The physical properties of
JU3 were investigated from earthbound orbit under favorable observational conditions
in  2007--2008 and 2011--2012. It is classified as a C-type
asteroid on the basis of optical and near-infrared spectroscopic observations
\citep{Binzel2001,Vilas2008,Lazzaro2013,Sugita2013,Pinilla2013,Moskovitz2013}.
Optical lightcurve observations indicated that JU3 is nearly spherical
with an axis ratio of 1.1--1.2 and a synodic rotational period of 7.625
$\pm$ 0.003 h \citep{Abe2008,Kim2013,Moskovitz2013}. Mid-infrared
photometry revealed that the asteroid has an effective diameter of $\sim$0.9 km
\citep{Hasegawa2008,Campins2009,Mueller2011}.

\citet{Mueller2014} recently constructed a sophisticated thermal model for JU3.
They determined an effective diameter of 870 $\pm$ 10 m
as well as the possible pole orientation and thermal inertia.
Motivated by their work, we thoroughly investigated the surface optical
properties of JU3 as part of the activity of the Hayabusa 2 project.
This paper thus aims to determine the geometric albedo, taking into account the updated effective diameter,
and characterize the optical properties useful for remote-sensing observations.
In particular, we emphasized the analysis of photometric data at low phase angles.
In Section 2, we describe the data acquisition and analysis. In Section 3, we employ
three photometric models to fit the magnitude--phase relation. Finally, we discuss our results
in Section 4 in terms of the mission target of Hayabusa 2.

\section{Data Acquisition and Analysis}

\subsection{Observations}
\begin{figure}
\epsscale{0.90}
\plotone{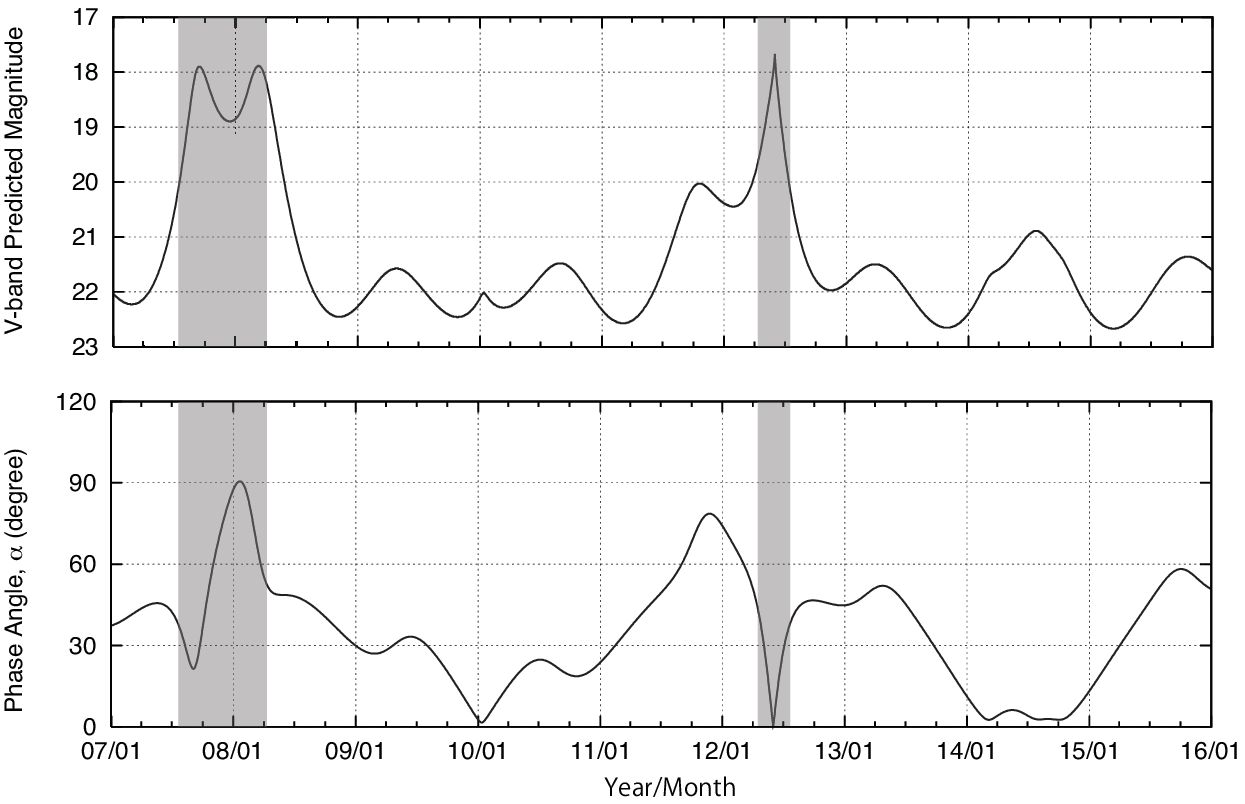}
\caption{Predicted V magnitude (top) and phase angle (bottom) of JU3 in 2007--2015. There were two observational
opportunities in late 2007--early 2008 and late 2011--mid-2012. Note that the magnitude is not the observed value.
Although these magnitudes have some degree of uncertainty, they provide a good perspective for understanding the observed
condition of the asteroid.
\label{fig:f1}}
\end{figure}

Figure \ref{fig:f1} shows the predicted V-band magnitudes
and phase angles for given dates in 2007--2016 obtained by NASA/JPL's Horizons
ephemeris generator\footnote{http://ssd.jpl.nasa.gov/}. The data set in 2007--2008 covers
a large phase angle  (up to 90\arcdeg), whereas that in 2011--2012 covers intermediate to
low phase angles (down to 0.3\arcdeg). The combined data thus provide a unique data set for studying
the magnitude--phase relation of the C-type asteroid in a wide phase
angle range in which main-belt asteroids cannot be observed from earthbound orbit.
A worldwide optical observation campaign for JU3 was conducted in 2011--2012
not only to support the Hayabusa 2 project but also to explore the physical nature of the target
asteroid. In particular, the campaign gave considerable weight to acquiring the relative
magnitudes for deriving the rotational period and shape model \citep{Kim2013,Kim2014}.
Among the campaign data, we selected magnitude data taken with calibration
standard stars under good weather conditions. In addition, we  chose data with a substantial signal-to-noise
ratio (i.e., S/N $>$ 10) and/or data with even phase angle coverage.
We also placed a priority on data at low phase angles. In total, we selected
data for JU3 taken on 21 nights with 8 telescopes.

The journal of observations is summarized in Table \ref{table:journal}, which includes observations
in \citet{Kawakami2009}, whose magnitude data were calibrated in a standard manner using photometric
standard stars but are not published in any journal.
The data we used in this work were taken with the following telescopes and instruments:
the University of Hawaii 2.2-m telescope (UH2.2m) with the Tektronix 2048 $\times$ 2048 pixel CCD
camera (Tek2k) atop Mauna Kea, USA; the Magellan I 6.5-m Baade telescope (Magellan-1) with the Inamori
Magellan Areal Camera and Spectrograph (IMACS); the Magellan II 6.5-m Clay telescope (Magellan-2)
with the Low Dispersion Survey Spectrograph (LDSS3) at the Las Campanas Observatory, Chile;
the ESO/MPI 2.2-m telescope (ESO/MPI) with the 7-Channel Imager, GROND, at $g'$, $r'$, $i'$, $z'$, $J$, $H$, and $K$
\citep{Greiner2008,Mueller2014} at the La Silla Observatory, Chile; the Nayuta 2.0-m telescope
(Nayuta) with the 2048 $\times$ 2064 pixel Multiband Imager (MINT) at the Nishi-Harima Astronomical
Observatory, Japan; the Indian Institute of Astrophysics 2.0-m Himalayan Chandra Telescope (HCT)
with the Himalayan Faint Object Spectrograph Camera (HFOSC)  at the Indian Astronomical Observatory, India;
the Tenagra II 0.81-m telescope (Tenagra II) with the 1K $\times$ 1K CCD camera (1KCCD) in Arizona, USA;
and the InfraRed Survey Facility 1.4-m telescope (IRSF) with
the Simultaneous three-color InfraRed Imager (SIRIUS) \citep{Nagayama2003} at the South African
Astronomical Observatory, South Africa. All but one of these telescopes were operated in a non-sidereal
tracking mode, whereas the HCT was operated in a sidereal tracking mode.
All the instruments were employed in an imaging mode with the standard broadband astronomical filters
listed in Table \ref{table:journal}. In addition to these new data taken in 2011--2012, we re-analyzed
some of the data taken in 2007 with the 1-m telescope at Lulin Observatory with the 1K $\times$ 1K CCD camera
(Taiwan) and  the Steward Observatory 1-m telescope in Arizona (USA).

\subsection{Pre-processing and Photometry}
The observed images were analyzed in the standard way for CCD imaging data. The raw data were subtracted
using bias frames (zero exposure) or dark frames (if the instruments were not cooled sufficiently to suppress
dark current) that were taken at intervals throughout each night. We used median-stacked data frames
using object frames to construct flat-field images with which to correct for vignetting of the optics and pixel-to-pixel
variation in the detector's response. The aperture photometry was conducted using the {\it apphot package}
in IRAF, which provides the magnitude within synthetic circular apertures projected onto the sky.
The parameters for the aperture photometry were determined manually depending on the image quality,
but in principle, we set an aperture radius of about 2--3 times the full width at half-maximum (FWHM),
which is large enough to enclose the detected flux of the asteroids and standard stars.
The sky background was determined within a concentric annulus having projected inner and
outer radii of $\sim$3$\times$FWHM and $\sim$4$\times$FWHM for point objects, respectively.
Flux calibration was conducted using standard stars in the Landolt catalog for the Johnson-Cousins $B$-, $V$-,
$R_\mathrm{C}$-, and $I_\mathrm{C}$-band filters \citep{Landolt1992,Landolt2009};
the Sloan Digital Sky Survey (SDSS) catalog for the $g'$-, $r'$-, $i'$-, and $z'$-band filters \citep{Ivezic2007};
and the Two Micron All Sky Survey (2MASS) catalog for the $J$-, $H$-, and $K$-band filters \citep{Skrutskie2006}.
At Nishi-Harima Astronomical Observatory, the data were  taken under variable sky conditions,
and the data taken with the Tenagra II and Magellan II were obtained without taking
standard star fields. To calibrate these data, we made a follow-up observation of the same sky field
with the UH2.2m and the same filter system.

\subsection{Derivation of Reduced $V$ Magnitude}
As described above, our data were obtained with a variety of filters. To derive the magnitude--phase relation
of JU3 using these data,  we first examined the color relations among the observed filter bands.
The  spectra of JU3 at 0.44--0.94~\micron\ has been studied well and shows no rotational variability
at the level of a few percent \citep{Moskovitz2013}.  We calculated the color indices using the spectrum in
\citet{Abe2008}, where they combined optical and near-infrared spectra taken at the MMT Observatory and
the NASAInfrared Telescope Facility \citep[see also][]{Vilas2008,Moskovitz2013}.
We considered that the observed asteroid spectrum is a product of the asteroid's reflectance and
the solar spectrum, and calculated the color indices (differences in magnitudes at two different bands)
using the following equation:

\begin{eqnarray}
\left(m_j-m_k\right) =
 -2.5\log\left(\frac{r_j}{r_k}\right)+\left(m_j-m_k\right)_\odot,
\label{eq:eq1}
\end{eqnarray}

\noindent
where $r_j$ and $r_k$ are the reflectances at the effective wavelengths of the $j$-th and $k$-th band filters, respectively,
and $\left(m_j-m_k\right)_\odot$ is the color index of the sun between the $j$-th and $k$-th bands.
We adopted the color indices of the sun and their uncertainties in \citet{Holmberg2006}.
We used the formula in \citet{Jordi2006} for converting from the Johnson-Cousins
$BVR_\mathrm{C}I_\mathrm{C}$ system to the SDSS $g'r'i'z'$ system.  Table \ref{tab:table2} compares
the observed and calculated color indices.
The observed color indices except for $g'$-$J$ (taken with GROND) were found to match the calculated
color indices to within the accuracy of our measurements ($\sim$5\%).
We noticed that the $g'$ magnitude taken with GROND has an ambiguity in the calibration process
due to unstable weather; it seems that the calculated $g'$-$J$ color index could be more reliable
than the observed one. Hereafter, we adopted the calculated color indices for deriving the
corresponding $V$ magnitude and tolerated the uncertainty of 5\% associated with the photometric system
conversion in the following discussion.

In addition to the phase angle dependence, the observed magnitude of the asteroid changed in
time because of its rotation. For JU3, the rotational effect results in a magnitude change
no larger than $\sim$0.1 mag with respect to the mean magnitude because it is nearly spherical
with an axis ratio of 1.1--1.2 \citep{Kim2013}.
To determine the magnitude--phase relation, we corrected the rotational modulation and derived the
magnitude averaged over the rotational phase. If observations cover both the peak and the trough,
we can derive the mean magnitude as a representative value from a single-night observation.
However, because most of the data could not cover a substantial portion of the rotational phase, we may mistake
the mean magnitudes. Although the effect is not as large as a half-amplitude of the lightcurve
(i.e., 0.1 mag or less), we corrected the rotational effect in 2012 data by deriving the zeroth-order term from the nightly
data using an empirical Fourier model to describe the relative magnitude change due to the rotation of JU3.
It is given by the following fourth-order Fourier series \citep{Kim2014}:

\begin{eqnarray}
\Phi\left(t\right) = \sum_{h=1}^4 \left[
     A_{2h-1} \sin\left(2\pi fh(t_{JD}-t_0)\right)
  +  A_{2h}   \cos\left(2\pi fh(t_{JD}-t_0)\right)
					\right] ~~,
\label{eq:eq2}
\end{eqnarray}

\noindent
where $\Phi(t)$ denotes the relative magnitude change caused by the rotation of JU3,
$f$ = 3.1475 is the rotational frequency in day$^{-1}$, and $t_{JD}$ and $t_0=$ 2456106.834045
(08:01:01.49 UT on 2012 June 28) are the observed median time in Julian days and offset time for
phase zero \citep[see Figure 1 in][]{Kim2013}, respectively. $A_{2h-1}$ and $A_{2h}$ are constants given by
$A_1$ = 0.0067, $A_2$ = 0.030, $A_3$ = 0.010, $A_4$ = 0.055, $A_5$ = 0.031, $A_6$ = 0.019,
$A_7$ = 0.0070, and $A_8$ = 0.0033. Assuming the observed magnitudes are given by
$\Phi(t)+H_\mathrm{V}(\alpha)$, we fitted the observed magnitudes after color correction
using Eq. (\ref{eq:eq2}) and derived the inferred mean magnitudes, $H_\mathrm{V}(\alpha)$.
To evaluate how well the empirical lightcurve model reproduces the
observed lightcurve, we compared the model in Eq. (\ref{eq:eq2}) with lightcurve data taken at several
different nights in 2012; these include light curves taken on 2012 May 31 with MOA-cam3 attached to MOA-II telescope
at Mt. John Observatory, New Zealand \citep{Sako2008,Sumi2010}, and on 2012 July 17--19 with HCT.
We found that the deviation is no larger than a few percent. We considered
a 4\% uncertainty associated with the correction for the lightcurve.
Regarding 2007--2008 data, the lightcurve model may not applicable because of the long interval
of time between $t_{JD}$ and $t_0$. We adopted the observed mean magnitudes 
and considered a 10\% uncertainty.

The observed corresponding $V$ magnitude, $H_\mathrm{V}(\alpha)$, was converted into
the reduced V magnitude $H_\mathrm{V}(1,1,\alpha)$, a magnitude at a hypothetical position
in the solar system, that is, a heliocentric distance $r_h$ = 1 AU, an observer's
distance of $\Delta$ = 1 AU, and a solar phase angle $\alpha$, which is given by

\begin{eqnarray}
H_\mathrm{V}(1,1,\alpha)=H_\mathrm{V}(r_h, \Delta, \alpha) - 5~\log(r_h \Delta)~~.
\label{eq:REDUCED}
\end{eqnarray}


\section{Results}

\begin{figure}
\epsscale{0.9}
\plotone{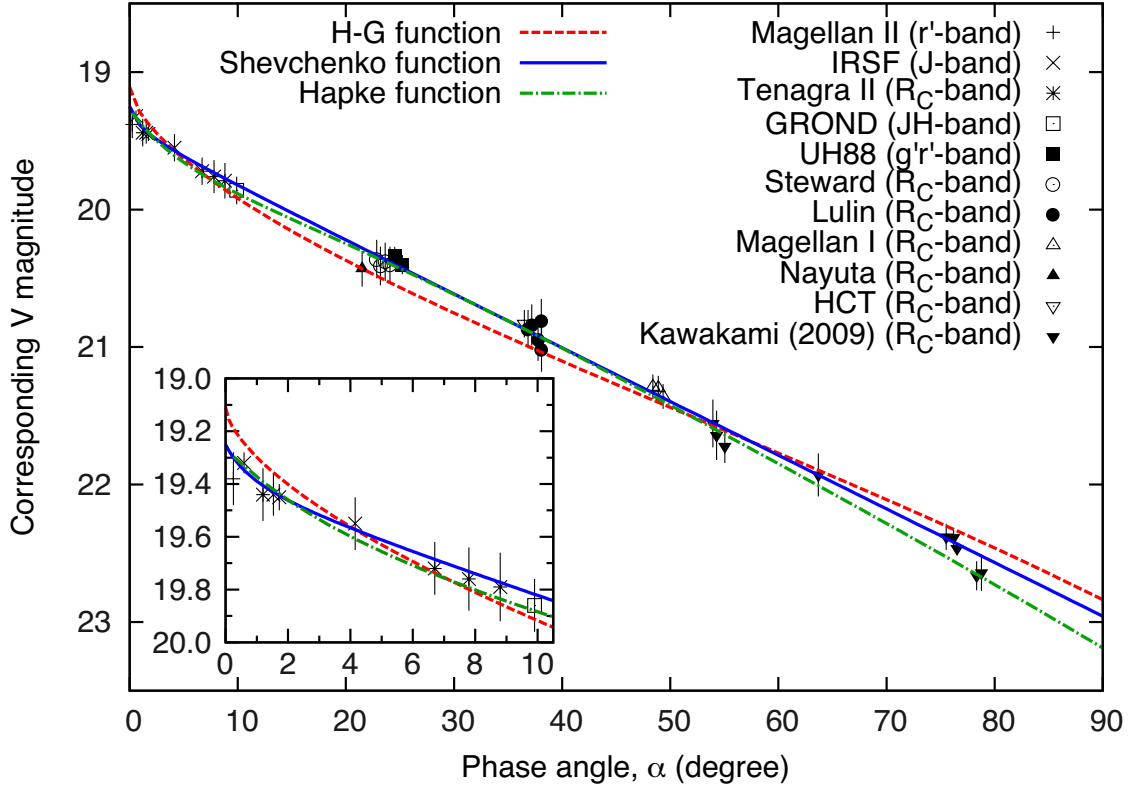}
\caption{Magnitude--phase relation of JU3. All observed magnitudes were converted into corresponding V
magnitudes assuming the color of the asteroid in Table \ref{tab:table2} at a hypothetical position of 1 AU from
the sun and observer. We show three models, that is, IAU $H$--$G$ function with our best fit parameters,
$H_\mathrm{V}$ = 19.13 $\pm$ 0.03 and $G$ = $-0.006 \pm$ 0.012, Shevchenko function with the best fit parameters,
$a$ = 0.21 $\pm$ 0.04, $b$ = 0.039 $\pm$ 0.001, and $H_\mathrm{V}$ = 19.24 $\pm$ 0.03, and Hapke model
with one of the best fit parameter sets, $\bar{\theta}=20$\arcdeg, $w=0.041$, $g=-0.38$, $B_0=1.43$, and
$h=0.050$.
\label{fig:f2}}
\end{figure}

In Figure \ref{fig:f2}, we show the magnitude--phase relation.
It shows obvious phase darkening, which is commonly observed in small solar system bodies.
The photometric calibration and color index correction appeared to work well, not only because
our new data set is smoothly connected to data from \citet{Kawakami2009}, but also
because there is no systematic displacement according to the observed filters.
In the following subsections, we apply three photometric functions to characterize
the magnitude--phase relation of JU3. For the fitting, we employed the Levenberg--Marquardt algorithm
to iteratively adjust the parameters to obtain the minimum value of $\chi^2$,
which is defined as

\begin{eqnarray}
\chi^2 =  \frac{1}{N} \sum^{N}_{n=1}\left[H_\mathrm{V}(1,1,\alpha)-\mathcal{H}_\mathrm{V}(1,1,\alpha)\right]^2,
\label{eq:hapke6}
\end{eqnarray}

\noindent where $\mathcal{H}_\mathrm{V}(1,1,\alpha)$ is the reduced magnitude calculated by each model.
$N$ is the number of magnitude data points. The observed data points are weighted by
$\sigma H_\mathrm{V}(1,1,\alpha)^{-2}$ for the fitting, where $\sigma H_\mathrm{V}(1,1,\alpha)$ denotes
the error of reduced magnitude.

\subsection{IAU $H$--$G$ formalism fitting}

We initially applied the $H$--$G$ formalism described in \citet{lumme1984} and \citet{Bowell1989}.
The function was adopted by the International Astronomical Union (IAU) and has been widely used
as a standard asteroid phase curve. It has the mathematical form

\begin{eqnarray}
\mathcal{H}_\mathrm{V}(1,1,\alpha) = H_\mathrm{V} - 2.5 \log \left[ \left(1-G\right) \exp \left(-3.33
				      \tan^{0.63} (\alpha / 2) \right) +
G \exp \left( -1.87 \tan^{1.22} (\alpha / 2) \right)\right] ~,
\label{eq:HG}
\end{eqnarray}

\noindent
where $H_\mathrm{V}$ is the absolute magnitude in the V band, and $G$ is
the slope parameter indicative of the steepness of the phase curve.
By fitting the entire data set, we obtained values of $H_\mathrm{V}$ = 19.12 $\pm$ 0.03
and $G$ = $-0.03$ $\pm$ 0.01 (dashed line in Figure \ref{fig:f2}). 
A careful examination of the fitting results reveals that
two-parameter fitting with the $H$--$G$ formalism cannot fit the entire magnitude--phase
relation, causing a discrepancy in the absolute magnitude.
The fitting does not match the  observational magnitude at small phase angles ($\alpha$ $<$ 2\arcdeg). 
In addition, the obtained slope parameter $G$ takes a peculiar negative value, although asteroids usually
take positive values \citep{harris1989}. Because the $H$--$G$ formalism has been usually applied
for main-belt asteroids observed at $\alpha \lesssim 30$, we fitted the data at $\alpha$ $<$ 30\arcdeg.
We obtained values of $H_\mathrm{V}$ = 19.22 $\pm$ 0.02 and $G$ = 0.13 $\pm$ 0.02.
We obtained the moderate $G$ value this time, although the model is largely deviated from the observed
data at $\alpha>$40\arcdeg.

\subsection{Shevchenko function fitting}

A simple but judicious model was proposed by \citet{Shevchenko1996}
for approximating an asteroid's magnitude--phase curves. It has the form

\begin{eqnarray}
\mathcal{H}_\mathrm{V}(1,1,\alpha)=C -\frac{a}{1+\alpha} + b~ \alpha ~~,
\label{eq:eq3}
\end{eqnarray}

\noindent
where $a$ is a parameter that characterizes the opposition effect amplitude, and
$b$ is a slope (mag deg$^{-1}$) describing the linear part of
the phase dependence. $C$ is given by $C$ = $H_\mathrm{V}+a$.
By fitting the entire data set, we obtained
$H_\mathrm{V}$ = 19.24 $\pm$ 0.03, $a$ = 0.19 $\pm$ 0.04, and $b$ = 0.039 $\pm$ 0.001 mag deg$^{-1}$.
It is interesting to notice that the Shevchenko function fits the observed data in a wide phase angle range
from 0.3\arcdeg\ to 75\arcdeg, although it was contrived to fit observed data at small phase angles (i.e.
$\alpha<$40\arcdeg). We obtained an absolute magnitude close to
that obtained by the $H$--$G$ formalism for the data at $\alpha<$30\arcdeg.

\begin{figure}
\epsscale{0.6}
\plotone{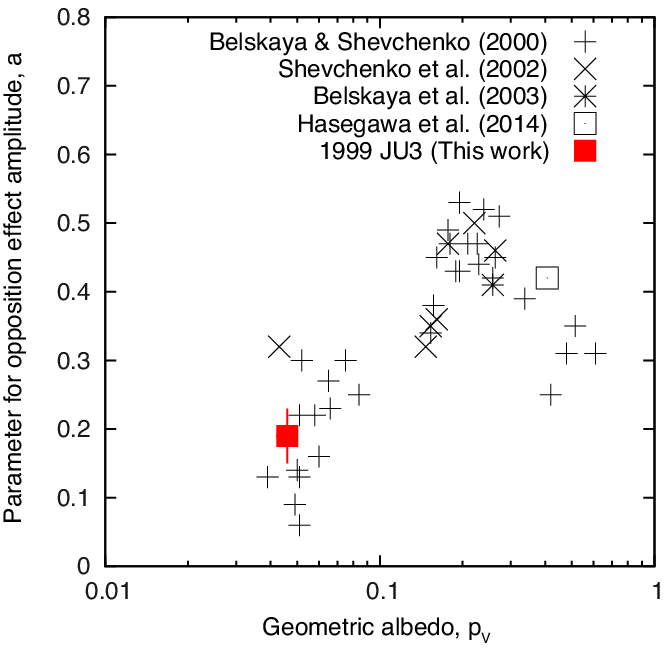}
\plotone{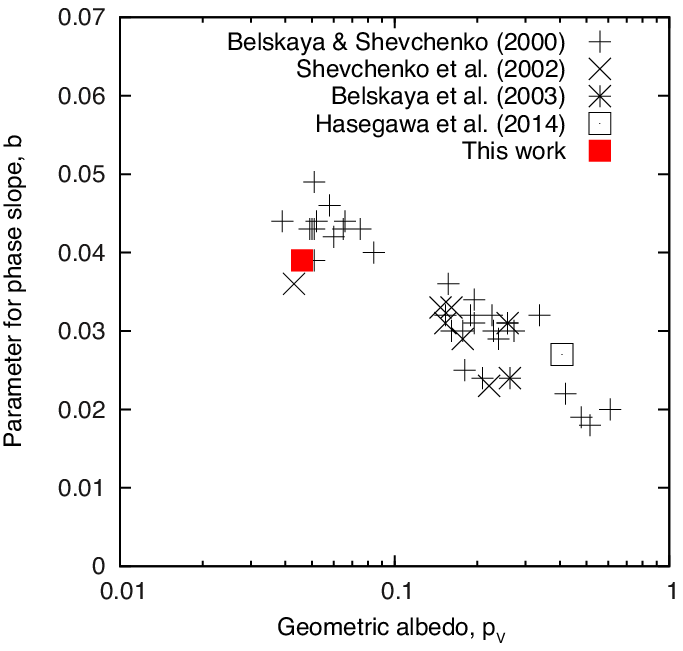}
\caption{Geometric albedo dependences of parameters for the opposition effect amplitude and linear phase slope
in Shevchenko's function. The original data were obtained from \citet{belskaya2000}, \citet{Shevchenko2002}, 
\citet{Belskaya2003}, and \citet{Hasegawa2014}. We updated the albedo values using data in \citet{Usui2011}.
\label{fig:belskaya}}
\end{figure}

The Shevchenko function has an  advantage in that it provides a direct
derivation of the opposition effect amplitude. With the best fit parameter of $a$ and
$b$, we computed the reflectance ratio of $I(0.3\arcdeg)$/$I(5\arcdeg)$ = 1.31 $\pm$ 0.05,
where $I(0.3\arcdeg)$ and $I(5\arcdeg)$ are the reflectances at $\alpha$ = 0.3\arcdeg\
and $\alpha$ = 5\arcdeg, respectively.
Figure \ref{fig:belskaya} shows the albedo dependence of parameters for the opposition effect amplitude
and linear phase slope of JU3, which are compared with those of $\gtrsim$10-km asteroids
\citep{belskaya2000,Shevchenko2002,Belskaya2003,Hasegawa2014}.
As noted in \citet{belskaya2000}, the phase slope
increases linearly as the albedo decreases.
They provided a phenomenological model to relate the phase slope
and the geometric albedo. Applying their empirical law, we obtained a
geometric albedo of 0.08 $\pm$ 0.03. The albedo is in accordance
with the geometric albedo of JU3 (see below).
The consistency of the phase slopes and the opposition effect amplitudes
between $\gtrsim$10-km asteroids and sub-km asteroid JU3 may suggest
that the magnitude--phase relation would be dominated by the albedo, not
by the asteroid diameter.
The amplitude and width of the opposition effect are considered to represent
the distance from the sun theoretically \citep{Deau2012}. The effect is caused by the apparent
size  of the sun when viewed from asteroids. JU3 was observed at a solar
distance of $r_h$ = 1.36 AU, whereas the other asteroids in Figure \ref{fig:belskaya}
were observed at $r_h$ = 1.8--3.4 AU. No significant difference is found between JU3 and the
other dark asteroids, most likely because the error is not small enough to detect
such a solar distance effect.

\subsection{Hapke model fitting}

Finally, we employed the Hapke model \citep{hapke1981,hapke1984,hapke1986}.
It has been applied to in-situ observational data taken with spacecraft onboard cameras.
The Hapke model provides an excellent approximation of the photometric function to correct for different
illumination conditions. However, because it has a complicated mathematical
form with many parameters (typically five or six), it often does not give a unique
fit to the limited observational data \citep{Helfenstein1989,belskaya2000}. Some parameters
are correlated with others to compensate one another, so there are a number of
best-fit parameter sets. Despite this complexity, the application of the Hapke model
is attractive in preparation for in-situ observation by Hayabusa 2.

The observed corresponding V magnitudes are converted
into the logarithm of $I/F$ (where $F$ is the incidence solar irradiance divided by $\pi$, and $I$ is the intensity of
reflected light from the asteroid surface) as

\begin{eqnarray}
-2.5 \log \left(\frac{I}{F}\right) = \mathcal{H}_\mathrm{V}(1,1,\alpha) - m_{V \odot} -\frac{5}{2} \log
\left( \frac{\pi}{S}\right) + m_c ~~,
\label{eq:hapke1}
\end{eqnarray}

\noindent
where $m_{V \odot}$ is the V-band magnitude of the sun at 1 AU,
$S$ is the geometrical cross section of JU3 in m$^2$, and
$m_c$ = $-5\log (1.4960 \times 10^{11})$ = $-55.87$ is a constant
to adjust the length unit. We used $m_{V \odot}$ = $-26.74$ \citep{Allen1973}.
We took the apparent cross section of JU3
$S$ = (5.94 $\pm$ 0.27) $\times$ 10$^5$ m$^2$, which is the equivalent area of a circle with
a diameter of 870 $\pm$ 10 m \citep{Mueller2014}.
The original Hapke model was contrived to characterize the bidirectional reflectance
of airless bodies \citep{hapke1981}. However, the observed quantity in this paper is the magnitude
integrated over the sunlit hemisphere observable from ground-based observatories.
We adopted the Hapke function integrating the intensity per unit area over that portion of
a spherical body \citep{hapke1984}. The equation is given as

\begin{eqnarray}
\nonumber
\frac{I}{F} =
 \left[\left(\frac{w}{8}\left[\left(1+B(\alpha)\right)P(\alpha)-1\right]+\frac{r_0}{2}(1-r_0)\right)
  \left(1-\sin\left(\frac{\alpha}{2}\right)\tan\left(\frac{\alpha}{2}\right)
   \ln \left[\cot\left(\frac{\alpha}{4}\right)\right]\right) \right.\\
   \left. +\frac{2}{3}r_0^2
   \left(\frac{\sin(\alpha)+(\pi-\alpha)\cos(\alpha)}{\pi}\right)\right]K(\alpha,\bar{\theta})~~,
\label{eq:hapke2}
\end{eqnarray}

\noindent
where $w$ is the single-particle scattering albedo. $K(\alpha,\bar{\theta})$ is a function that corrects
for the surface roughness parameterized by $\bar{\theta}$ \citep{hapke1984}.
The term $r_0$ is given by

\begin{eqnarray}
r_0 = \frac{1-\sqrt{1-w}}{1+\sqrt{1-w}}~~.
\label{eq:hapke3}
\end{eqnarray}

The opposition effect term $B(\alpha)$ is given by

\begin{eqnarray}
B(\alpha) = \frac{B_0}{1+\frac{\tan(\alpha / 2)}{h}}~~,
\label{eq:hapke4}
\end{eqnarray}

\noindent
where $B_0$ characterizes the amplitude of the opposition effect, and $h$
characterizes the width of the opposition effect. We used the
one-term Henyey--Greenstein single particle phase function solution
\citep{henyey1941}:

\begin{eqnarray}
P(\alpha)=\frac{(1-g^2)}{(1+2g \cos (\alpha) + g^2)^{3/2}}~~,
\label{eq:hapke5}
\end{eqnarray}

\noindent
where $g$ is called the asymmetry factor. Positive values of $g$
indicate forward scatter, $g$=0 isotropic, and negative $g$ backward scatter.

\begin{figure}
\epsscale{0.9}
\plotone{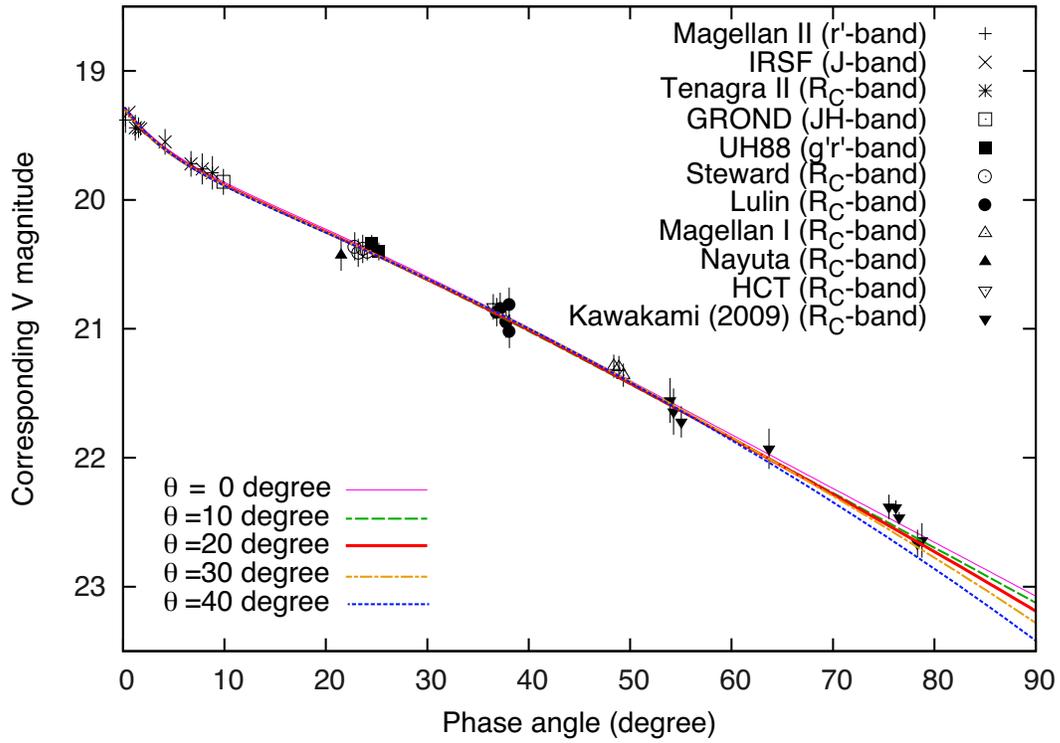}
\caption{Comparison between five sets of Hapke model parameters in Table 3 ($\bar{\theta}=$0--40\arcdeg\ from top to bottom). 
\label{fig:hapke}}
\end{figure}

For the fitting, we considered initial values in the possible ranges of the parameters, that is,
$0.01\leq w\leq 0.09$ at intervals of 0.02, $0.01\leq h\leq 0.1$ at intervals of 0.03, $-0.5\leq g\leq -0.1$
at intervals of 0.2, $0\leq B_0\leq 4.0$ at intervals of 1, and $0\arcdeg\leq \bar{\theta} \leq 40\arcdeg$
at intervals of 10\arcdeg, and conducted the parameter fitting. However, we soon realized that
the fitting algorithm cannot converge when these five parameters are variables.
We found that, mathematically, the effect of the surface roughness, $\bar{\theta}$,  can be compensated
by the other parameters to produce nearly identical disk-integrated curves; consequently,
the fitting algorithm cannot converge. A similar argument was made for Itokawa's disk-integrated
function, where it was claimed that high phase angle data
are necessary to constrain the surface roughness effect \citep{lederer2008}.
We assumed $\bar{\theta}$ to be 0\arcdeg, 10\arcdeg,
20\arcdeg, 30\arcdeg, or 40\arcdeg\ and derived the other parameters for the given $\bar{\theta}$ values.
The obtained Hapke parameters are summarized in Table \ref{table:hapke}
and the magnitude--phase relation with the parameters is shown in Figure \ref{fig:f2}
and Figure \ref{fig:hapke}. In terms of $\chi^2$, there is no significant difference between
the models with different $\bar{\theta}$. However, our magnitude data at high phase angles favour
models with $\bar{\theta}\lesssim$30\arcdeg (see Figure \ref{fig:hapke} at $\alpha>$70\arcdeg).

In Table \ref{table:hapke}, we calculated the Bond albedo (also known as the spherical albedo),
$A_B$, which is the fraction of incident light scattered in all
directions by the surface. It is given by $A_B$ = $q p_\mathrm{V}$, where $q$ is
the value of the phase integral, defined as

\begin{eqnarray}
q = 2 \int_0^{\pi} \frac{\Phi(\alpha)}{\Phi(0)}~\sin(\alpha)~d\alpha~~.
\label{eq:phaseintegral}
\end{eqnarray}

In comparison with the other mission target asteroids,
JU3 has Hapke parameters similar to those of the C-type asteroid Mathilde.
We noticed that the resultant albedos are independent of the surface roughness,
although the other Hapke parameters depend on the assumed roughness.
The Hapke model matches well Shevchenko's model at the lowest phase
angle, yielding $H_\mathrm{V}$ = 19.24 $\pm$ 0.03. From the absolute magnitude,
we can conclude that JU3 has a geometric albedo of $p_\mathrm{V}$ = 0.046 $\pm$ 0.004.
The derived albedo is consistent with those of C-type asteroids, that is, 0.071$\pm$0.040
\citep{Usui2013}.

\section{Discussion}

\subsection{Error Analysis}

As we described above, we considered possible error sources with conservative estimates and obtained
the total error. Eventually, each corresponding V magnitude has a photometric error of $\sim$0.1 mag.
The photometric models above fit the observed data within the photometric errors, most likely because we
may give adequate consideration to the errors. As a result, each error in the magnitude data taken with
a variety of filters and instruments seems to be randomized, which provides a reliable absolute magnitude.
We obtained a fitting error of the absolute magnitude of 0.03 mag. 
It is important to note that the absolute magnitude we derived differs greatly (0.4 mag) from that in the previous
study \citep{Kawakami2009}, which was obtained on the basis of observation at $\alpha >$ 22\arcdeg\
and is commonly referred to in previous research to derive the geometric albedo
\citep{Hasegawa2008,Campins2009,Mueller2011}. 
In general, the absolute magnitudes of main-belt asteroids have been determined through ground-based
observations at $\alpha<30$\arcdeg. It is likely that \citet{Kawakami2009} might lead inadequate result
in the absolute magnitude due to the paucity of data at low phase angles.
Because our data set covers low phase angles almost equivalent to half of the apparent solar disk size,
it provides a more reliable absolute magnitude without extrapolating the phase function.

A drawback of our measurement is that we ignored the wavelength dependence of the magnitude--phase
relation. In fact, the single scattering albedo ($w$) and opposition surge width ($h$) of S-type asteroids
Itokawa and Eros exhibit strong wavelength dependences \citep{Kitazato2008,clark1999}.
Figure \ref{fig:f2} shows the magnitude--phase relation for the various filters. In particular, we have $J$-band
and $R_\mathrm{C}$-band data at $\alpha<$ 10\arcdeg. Because the phase slopes of these two bands are
nearly the same, we conjecture that the magnitude--phase relation is less dependent on the wavelength.
In general, the observed magnitude--phase relation can be influenced by
a combination of the shadow-hiding effect and coherent backscattering mechanism, and these effects
depend on the geometric albedo \citep{belskaya2000}. Unlike Itokawa and Eros, which have very red spectra
with moderate absorption around 1 \micron, JU3 has almost constant reflectance from
0.45 \micron\ to $\sim$1.5 \micron\  \citep{Moskovitz2013}. Accordingly, we may attribute the weak
wavelength dependence of the phase curve to the flat spectrum of JU3.

Regarding the errors of the Hapke parameters, a large uncertainty remains because of the undetermined $\bar{\theta}$.
The $\chi^2$ derived by Shevchenko's function is equivalent to that derived by the Hapke function, suggesting
that three-parameter fitting for Shevchenko's function is mathematically sufficient to characterize the observed
magnitude--phase relation. We fixed the surface roughness parameter $\bar{\theta}$ and derived the other
four parameters in the Hapke model. It seems that
$B$ and $h$ characterize the shape and amplitude of the opposition effect, which correspond to a parameter,
$a$, in Shevchenko's function, whereas the other two parameters adjust the vertical magnitude offset and the
entire phase slope, which correspond to $H_\mathrm{V}$ and $b$ in Shevchenko's function.
Among the Hapke parameters, we determined $h$ independently of $\bar{\theta}$. 
$h$ can be interpreted in terms of porosity and grain size of the optically active regolith. 
Applying a model in \citet{hapke1986} and assuming a lunar-like grain size distribution
with a ratio of the largest particle size to the smallest particle size of 10$^3$, we obtained
the porosity of 0.64. The derived porosity suggests that the surface of JU3 may be covered with coarse regoliths
($\approx$1 cm, using an empirical size--porosity relation in \citealt*{Kiuchi2014}).

\subsection{In Preparation for the Hayabusa 2 Mission}
The primary goal of the Hayabusa 2 mission is to bring back samples from a C-type asteroid
or an asteroid analogous to C-type asteroids, following up on the Hayabusa 1 mission, which returned a
sample from the S-type asteroid Itokawa  \citep{Yoshikawa2008,Yoshikawa2012}.
Because of the similarity of the spectra, C-type asteroids are considered to be parent bodies of
carbonaceous chondrite meteorites, which are abundant in organic materials and hydrous minerals.
In the nominal plan, multiple samplings are scheduled to obtain different types of asteroid material
on the basis of their composition and degrees of aqueous alteration and space weathering.
Because the reflectance at 0.55 \micron\ depends mainly on the abundance of
carbon compounds, the surface reflectance map will provide information useful
for the selection of the sampling sites. Therefore, speedy  construction of the reflectance map
is crucial to the project's success.
The observed reflectances at given wavelengths should be converted to a standard geometry consisting of
the incident angle $i$ = 30\arcdeg\, emission angle $e$ = 0\arcdeg, and phase angle $\alpha$ = 30\arcdeg\
using a photometric model. Among the models, the Hapke model has been widely used to correct data
taken with spacecraft onboard cameras  \citep[see, e.g.,][]{clark1999,clark2002}.
Through this work, we have determined the Hapke parameters except for the surface roughness.
Because the disk-resolved reflectance is sensitive to the surface roughness, we expect that it will be obtained
uniquely once the JU3 images are taken. We thus propose to determine the surface roughness
while fixing the other four parameters on a restricted basis by the magnitude--phase relation.
The construction of JU3's shape model is essential to calculating the incident and emission angles.
Once the shape is determined, we expect that the full set of Hapke parameters will be fixed using the images
taken with the onboard cameras.


Our work provides useful information for the calibration of the onboard remote-sensing devices
such as the Optical Navigation Camera (ONC) and the Deployable CAMera (DCAM). Specifically, we provide
accurate magnitude models. Similar to the procedure on the Hayabusa 1 mission,
measurements  of JU3's lightcurve are planned for the purpose of flux calibration \citep{Sugita2013}.
A comparison of the magnitude--phase relation in this paper and the observed disk-integrated data count
will enable the determination of the calibration factors, as we did using stellar observations during the cruising phase
for Hayabusa Asteroid Multi-band Imaging Camera (AMICA) \citep{Ishiguro2010}. Unlike the case of AMICA
for Hayabusa 1, we expect that the calibration parameters will be determined easily thanks to the accurate
photometric models in this paper.

%

\section{Summary}

We examined the magnitude--phase relation of (162173) 1999 JU3
using data taken from ground-based observatories.
The major findings of this paper are as follows:

\begin{enumerate}
\item{
The IAU $H$--$G$ formalism does not fit the observed data at small and large phase
angles ($\alpha$ $<$ 2\arcdeg\ and $\alpha$ $>$ 50\arcdeg) simultaneously  using a single set of parameters.
By fitting the data set at $\alpha$ $<$ 30\arcdeg, we obtained values of $H_\mathrm{V}$ = 19.22 $\pm$ 0.02
and $G$ = 0.13 $\pm$ 0.02.
}

\item{
Using the Shevchenko function, we found that the magnitude--phase relation exhibits linear behavior
in a wide range of phase angles ($\alpha$ = 5--75\arcdeg) and shows weak nonlinear
opposition brightening at $\alpha<$ 5\arcdeg. The phase slope (0.039 $\pm$ 0.001 mag deg$^{-1}$)
and opposition effect amplitude (parameterized by the ratio of intensity at $\alpha$=0.3\arcdeg\
to that at $\alpha$=5\arcdeg, $I(0.3\arcdeg)$/$I(5\arcdeg)$=1.31$\pm$0.05)
are consistent with those of typical C-type asteroids.
}

\item{
We were not able to uniquely constrain the full set of parameters in the Hapke model, but were able
to find the best-fit values for four parameters given a range of fixed values for the surface roughness parameter.
Assuming $\bar{\theta}$=20\arcdeg (the typical value for C-type asteroids, \citet{clark1999}), we obtained
$w$=0.041, $g$=-0.38, $B_0$=1.43, and $h$=0.050. The results obtained are consistent with those of
the C-type asteroid Mathilde.
}

\item{
By combining this study with a thermal model of asteroids (M{\"u}ller et al.
in preparation), we obtained a geometric albedo of $p_\mathrm{v}$ = 0.047 $\pm$ 0.003, a phase
integral of $q$ = 0.32 $\pm$ 0.03, and a Bond albedo of $A_\mathrm{B}$ = 0.014 $\pm$ 0.002.
}

\end{enumerate}

\vspace{1cm}
{\bf Acknowledgments}\\
This research was supported by the Korea Research Council of Fundamental Science \&
Technology and by the Korea Astronomy and Space Science Institute. 
We would like to thank the referee, Dr. Bruce Hapke, for carefully reading our manuscript
and for giving  constructive comments. This paper was written
when the first author was at UCLA and Paris Observatory, where he was supported by Prof. David Jewitt
and Dr. Jeremie Vaubaillon. SH was supported by the Space Plasma Laboratory, ISAS, JAXA.
Part of the funding for GROND (both hardware as well as personnel) was generously granted from
the Leibniz-Prize to Prof. G. Hasinger (DFG grant HA 1850/28-1). P.S. acknowledges support
by DFG grant SA 2001/1-1.

\clearpage

\clearpage



\clearpage

\clearpage

\clearpage
\begin{deluxetable}{lccr}
\tabletypesize{\footnotesize}
  \tablecaption{Observation Summary}
\tablewidth{0pt}
\tablehead{
\colhead{Telescope}   & \colhead{Instrument} & \colhead{Filter} &
\colhead{Date (UT)} }
\startdata
{\it This work}\\
UH2.2m\tablenotemark{a} & Tek2048 & $r'$, $g'$ & 2012 June 26, 27, July 12\\
IRSF\tablenotemark{b} &  SIRIUS & $J$, $H$, $K$ & 2012 May 30, 31, June 1, 2, 4 \\
Magellan I\tablenotemark{c} & IMACS & $R_\mathrm{C}$ & 2012 April 5--7\\
Magellan II\tablenotemark{d} & LDSS3 & $g'$, $r'$, $i'$ & 2012 June 1\\
ESO/MPI\tablenotemark{e} & GROND & $g'$, $r'$, $i'$, $z'$, $J$, $H$, $K$ & 2012 May 28,
	     June 9--10\\
Tenagra II\tablenotemark{f} & SITe 1k CCD & $R_\mathrm{C}$ & 2012 May 31, June 7--9\\
Nayuta\tablenotemark{g} & MINT & $R_\mathrm{C}$ & 2012 June 22\\
HCT\tablenotemark{h} & HFOSC & $R_\mathrm{C}$ & 2012 July 17\\
\\
Steward\tablenotemark{i} & Optical CCD & $V$, $R_\mathrm{C}$ & 2007 September 10--13\\
Lulin\tablenotemark{j} & EEV 1k CCD   & $B$, $V$, $R_\mathrm{C}$, $I_\mathrm{C}$ & 2007 July 19--23, December 3, 7--8\\
        &       &            & 2008 February 27--28, April 2, 4--5 \\
\\
{\it Kawakami (2009)}\\
Kiso\tablenotemark{k} & 2KCCD & $B$, $V$, $R_\mathrm{C}$ & 2007 November 8\\
\enddata
\tablenotetext{a}{~The University of Hawaii 2.2-m Telescope, USA}
\tablenotetext{b}{~The InfraRed Survey Facility 1.4-m Telescope, South Africa}
\tablenotetext{c}{~Las Campanas Observatory Magellan I Baade Telescope, Chile}
\tablenotetext{d}{~Las Campanas Observatory Magellan II Clay Telescope, Chile}
\tablenotetext{e}{~La Silla Observatory ESO/MPI 2.2-m Telescope, Chile}
\tablenotetext{f}{~Tenagra II 0.81-m Telescope, USA}
\tablenotetext{g}{~The Nishi-Harima Astronomical Observatory Nayuta 2-m Telescope, Japan}
\tablenotetext{h}{~The Indian Institute of Astrophysics 2-m Himalayan Chandra Telescope, India}
\tablenotetext{i}{~Steward Observatory 61-in. (1.54-m) Kuiper Telescope, USA}
\tablenotetext{j}{~Lulin Observatory One-meter Telescope, Taiwan}
\tablenotetext{k}{~The University of Tokyo Kiso Observatory 1.05-m Schmidt Telescope, Japan}
\label{table:journal}
\end{deluxetable}

\clearpage
\begin{table}
\scriptsize
\begin{center}
\caption{Color Indices\label{tab:table2}}
\begin{tabular}{lcccccccc}
\tableline\tableline
 & $B$-$V$ & $V$-$R_\mathrm{C}$ & $V$-$I_\mathrm{C}$ & $g'$-$r'$ & $r'$-$i'$ & $V$-$J$ & $g'$-$J$ \\
\tableline
Observed\citep{Kawakami2009} & 0.66 (0.06) & 0.40 (0.06) & 0.74 (0.07) & -- & -- & -- \\
Observed (this work) &  & 0.37 (0.03) & -- & 0.47 (0.03) & -- & -- & 1.16 (0.05) \\
Calculated & 0.65 (0.01) & 0.34 (0.01) & 0.72 (0.01) & 0.45 (0.02) & 0.13
		     (0.01) & 1.21 (0.04) & 1.48 (0.05) \\
\tableline
\tablecomments{Numbers in the parentheses are errors of the color indices.}
\end{tabular}
\end{center}
\end{table}

\clearpage
\begin{deluxetable}{lcrrrrrrrrr}
\tabletypesize{\scriptsize}
  \tablecaption{\footnotesize{Hapke Parameters for JU3 and Other Mission Target Asteroids in V Band (550 nm)} }
\tablewidth{0pt}
\tablehead{
\colhead{ }   & \colhead{Type$^a$} & \colhead{$w^b$} & \colhead{$g^c$} &
 \colhead{$\bar{\theta}^d$ (\arcdeg)} & \colhead{$B_0^e$} & \colhead{$h^f$} &
 \colhead{$p_\mathrm{V}^g$} & \colhead{$q^h$} &
 \colhead{$A_\mathrm{B}^i$} & \colhead{Reference}
}

\startdata
\underline{This work}\\ 
     1999 JU3 & C$_\mathrm{g}$ & 0.038 & -0.39 &   [0] & 1.52 & 0.049 & 0.045 & 0.32 & 0.014  & \nodata\\ 
      &  & 0.039 & -0.39 & [10] & 1.49 & 0.049 & 0.045 & 0.32 & 0.014  & \nodata\\ 
      &  & 0.041 & -0.38 & [20] & 1.43 & 0.050 & 0.045 & 0.31 & 0.014  & \nodata\\
      &  & 0.046 & -0.36 & [30] & 1.34 & 0.051 & 0.045 & 0.31 & 0.014  & \nodata\\
      &  & 0.052 & -0.34 & [40] & 1.21 & 0.051 & 0.044 & 0.30 & 0.013  & \nodata\\
\\
\underline{Other mission targets}\\
     (253) Mathilde & C$_\mathrm{b}$ & 0.035 & -0.25 & 19 & 3.18 & 0.074 & 0.041 & 0.33 &
				     0.013 & \citet{clark1999} \\
     (243) Ida & S &0.22 & -0.33 & 18 & 1.53 & 0.02 & 0.21 & 0.34 & 0.070 &
					 \citet{helfenstein1996} \\
     (433) Eros & S & 0.43 & -- & 36 & 1.0 & 0.022 & 0.29 & 0.39 & 0.12
				     & \citet{domingue2002} \\
     (951) Gaspra & S & 0.36 & -0.18 & 29 & 1.63 & 0.06 & 0.22 & 0.49 & 0.11
				     & \citet{helfenstein1994} \\
     (25143) Itokawa & Sk & 0.70 & -- & 40 & 0.02 & 0.141 & 0.19 & 0.11 & 0.021 &
				     \citet{lederer2008}\\
     (5535) Annefrank & S & 0.63 & -0.09 & 49 & [1.32] & 0.015 &0.28 & 0.44 & 0.12 & \citet{Hillier2011}\\
     (4) Vesta & V & 0.51 & -0.24 & 18 & 1.83 & 0.048 & 0.42 & 0.47 & 0.20 & \citet{Li2013,Hasegawa2014}\\
     (2867) Steins & E & 0.66 & -0.30 & 28 & 0.60 & 0.027 & 0.39 & 0.59 & 0.24 & \citet{Spjuth2012}$^*$\\
\enddata
\tablecomments{
$^a$ Taxonomic type,
$^b$ Single-scattering albedo,
$^c$ Asymmetry factor,
$^d$ Roughness parameter,
$^e$ Opposition amplitude parameter,
$^f$ Opposition width parameter,
$^g$ Geometric albedo,
$^h$ Phase integral, and
$^i$ Bond albedo.
$^*$ Parameters obtained at 630 nm.
\\
Numbers in parentheses for $\bar{\theta}$ are fixed values.
}
\label{table:hapke}
\end{deluxetable}

\end{document}